  \documentclass[aps,a4paper,showpacs]{revtex4}
 \usepackage{amsmath}
\newcommand{\Label}[1]{\label{#1}}                  
\newcommand{\Bibitem}[1]{\bibitem{#1}}    
\newcommand{\be}{\begin{equation}}
\newcommand{\ee}{\end{equation}}
\newcommand{\ba}{\begin{eqnarray}}
\newcommand{\ea}{\end{eqnarray}}

\newcommand{\nn}{\nonumber\\}
\newcommand{\Ref}[1]{(\ref{#1})}
\newcommand{\av}[1]{\left \langle #1\right \rangle}
\newcommand{\half}{\textstyle{\frac{1}{2}}}
\newcommand{\br}{{\bf r}}
\newcommand{\bv}{{\bf v}}
\newcommand{\bk}{{\bf k}}
\newcommand{\bq}{{\bf q}}
\newcommand{\bj}{{\bf j}}
\newcommand{\bg}{{\bf g}}

\newcommand{\calO}{{\cal O}}

\newcommand{\calL}{{\cal L}}
\newcommand{\calP}{{\cal P}}

\newcommand{\calG}{{\cal G}}
\newcommand{\calLbar}{\overline{\cal L}}
\newcommand{\calLtilde}{\widetilde{\cal L}}

\newcommand{\calLinvert}{\calL^\epsilon}
\newcommand{\bF}{{\bf F}}
\newcommand{\bG}{{\bf G}}
\newcommand{\bR}{{\bf R}}

\newcommand{\bsigma}{{\boldsymbol\sigma}}

\begin{document}

\title{Generalized Green-Kubo formulas for fluids with\\
impulsive, dissipative, stochastic and conservative interactions}

\author{M.H. Ernst$^{1,2}$}
\author{R.\ Brito$^{2,3}$}
\affiliation{\it $\,^{1}$Institute for Theoretical Physics,
Universiteit  Utrecht, 3508 TD Utrecht, The Netherlands\\
\it $\,^{2}$Departamento~de F\'{\i}sica Aplicada I, Universidad Complutense, E--28040 Madrid, Spain  \\
 \it $\,^{3}$GISC, Universidad Complutense, E--28040 Madrid, Spain}

\date{\today}


\begin{abstract}

We present a generalization of the  Green-Kubo expressions for
thermal transport coefficients $\mu$ in complex fluids of the
generic form, $\mu= \mu_\infty +\protect{\int}_0^\infty dt V^{-1}
\langle J_\epsilon \exp(t {\cal L}) J \rangle_0$, i.e.\ a sum of
an instantaneous transport coefficient $\mu_\infty$, and a time
integral over a time correlation function in a state of thermal
equilibrium between a current $J$ and its conjugate current
$J_\epsilon$. The streaming operator $\exp(t{\cal L})$  generates
the trajectory of a dynamical variable $J(t) = \exp(t{\cal L}) J$
when used inside the thermal average $\langle \cdots\rangle_0$.
These formulas are valid for conservative, impulsive (hard
spheres), stochastic and dissipative forces (Langevin fluids),
provided the system approaches a thermal equilibrium state. In
general $\mu_\infty \neq 0$ and $J_\epsilon \neq J$, except for
the case of conservative forces, where the equality signs apply.
The most important application in the present paper is the hard
sphere fluid.

\end{abstract}

\pacs{{05.}{40.-a}, {05.}{20.Dd}}

\maketitle

\setcounter{equation}{0}
\section{Introduction}

As the interest in fluids during the last decades has been
shifting from standard fluids with smooth conservative
interactions to more complex fluids with ditto interactions, a new
analysis of linear response theory for such systems seems timely.
The present paper provides this analysis for a wide class of
complex fluid-type systems.

The Green-Kubo formulas for thermal transport coefficients in
simple classical fluids with {\it conservative} interactions are
widely used, and generally accepted
\cite{Hansen-McDonald,Allen-Tildesley,ED75,Nico} as exact expressions for
general densities, as long as the deviations from equilibrium and
the gradients are small, and the transport coefficients exist.
These expressions are given in terms of equilibrium time
correlation functions between $N$-particle currents, i.e.
\be\Label{11}
\mu = \int_0^\infty \! dt \lim_{V\to\infty}
\frac{1}{V} \langle J(0)J(t)\rangle_0,
\ee
where $\mu$ denotes a typical transport coefficient, $\langle
\dots \rangle_0$ is an average over a thermal equilibrium ensemble
at temperature $T=1/k_B\beta$, with a weight $\rho_0\sim
\exp[-\beta H]$, and $\lim_{V\to\infty}$ denotes the thermodynamic
limit. In the sequel this limit is understood, but not explicitly
written. The time evolution of a dynamical variable, $J(t)=
e^{t\calL} J(0)$, can be described by a streaming operator $
e^{t\calL}$, and is invariant under the time reversal
transformation. The infinitesimal generator, $\calL$, referred to
as Liouville operator,  changes sign under the time reversal
transformation. The total microscopic flux $J = \lim_{\bk \to 0}
j_\bk$ is related to the Fourier mode $a_\bk$ of a conserved
density through the local conservation law, $\partial_t a_\bk
\equiv \calL a_\bk = - i \bk \cdot \bj_\bk$, and contains in
general contributions from kinetic transport, and from collisional
transfer, i.e. transport through inter-particle forces. The simple
form \Ref{11} only applies to systems with time reversal invariant
dynamics, as we shall see in this paper.

A summary of the results for systems with non-conservative forces
has been given in Ref.\cite{EB-EPL}, and the goal of this article
is to provide a general analysis with detailed applications to
hard sphere fluids. The applications to stochastic systems, as
described by Langevin and Fokker-Planck equations will be given
elsewhere \cite{ME-Langevin}. Green-Kubo formulas still exist, but
their generic form is different from \Ref{11}. The systems of
interest are elastic hard sphere fluids with impulsive forces
\cite{DE04}, and  systems with Brownian dynamics \cite{Pias},
considered as a mixture of hard spheres where the mass ratio is
taken to infinity; lattice gas cellular automata \cite{DE-LGCA}
and multi-particle collision dynamics (MPCD) models, where the
discrete time dynamics involves stochastic variables
\cite{Kapral}. Finally there is the large class of complex fluids,
where $N$-particle fluid or magnetic systems are described by
mesoscopic Langevin equations containing dissipative and
stochastic forces, or by corresponding Fokker-Planck equations for
the  probability distribution
\cite{Koelman,pep+Warren,Marsh,PE95,RE05,etc,Hess,Rubi,Bonet,pep97,pep03,Marisol+EE,pep-Vasquez}.
In fact, also the models for critical and unstable fluids and
magnets \cite{RMP} belong to the Langevin models covered by the
present theory. We also note that the Langevin and Fokker-Planck
equations are not time reversal invariant. The theory to be
developed here may be applied as soon as the analytic form of the
generator $\calL$ and of the thermal distribution $ \rho_0$ are
given, i.e. as soon as the time evolution can be expressed either
in terms of hard sphere dynamics or in terms of Langevin and
Fokker-Planck equations.

The systems listed above not only refer to simple fluids, but
cover a large range of complex fluids, also outside the collection
of critical and unstable systems. For instance,
Ref.\cite{pep-Vasquez} considers diffusion coefficients of
colloidal particles, whereas Ref.\cite{Marisol+EE} contains an
extensive list of applications of the so-called dissipative 
particle dynamics (DPD) fluid to colloidal particles, emulsions and 
polymers \cite{Koelman,pep+Warren}.  These are just a few
examples, but there are many more non-equilibrium fluid-type
models, that take {\it fluctuations} into account, satisfy {\it
detailed balance} criteria, and reach {\it thermal equilibrium},
for which Green-Kubo formulas have been or can be derived.

Granular fluids are outside the  scope of this study, although
Green-Kubo formulas have been considered in the literature
\cite{TvN-Isaac,Dufty-Baskaran05}. The reason is that such systems
do not  approach a stable state of thermal equilibrium, which is a
basic requirement in the present derivation. In the above articles
the stable equilibrium state is replaced by the unstable
homogeneous cooling state. Such extensions are outside the scope
of the present paper.

The goal of  this study is to derive  generalized Green-Kubo
expressions for large classes of $N$-particle systems, that do
approach  thermal equilibrium, and have equations of motion that
possibly lack time reversal invariance. The interaction forces may
be impulsive, dissipative, stochastic, and possibly include
conservative forces as well. The method used is linear response
theory \cite{Hansen-McDonald}, which yield exact expressions for
transport coefficients, valid for arbitrary densities. They have
the generic form,
\be\Label{15} \mu=\mu_\infty
+V^{-1}\int_0^\infty\! dt
 \langle J_\epsilon e^{t\calL}J\rangle_0.
\ee
The {\it pseudo}-streaming operators $\exp(t\calL)$ with $t>0$ are
well-defined for {\it all} points in the relevant phase space, and
generate the proper trajectories of the dynamical variables when
used {\it inside statistical averages} $\av{\cdots}_0$. An
intuitive generalization of \Ref{11} to non-conservative forces
does not seems obvious. For the general class of Langevin systems
a result similar to \Ref{15} has been obtained in
Ref.\cite{pep-Vasquez}, but the instantaneous contribution
$\mu_\infty$ to the dissipative transport coefficient has not been
identified. The only known results of the complete form \Ref{15}
with instantaneous part as well as time correlation part have been
obtained in Ref.\cite{Pias} for the friction coefficient of
Brownian particles, and in Ref.\cite{DE04} for the shear viscosity
of the hard sphere fluid. In the former case the second term in
\Ref{15} contains a time correlation function of an inter-particle
force and its conjugate force, in the latter of a stress and a
conjugate stress.

A discussion about exact expressions for transport coefficients is
not complete without the Einstein-Helfand formulas for the
Navier-Stokes transport coefficients
\cite{Hansen-McDonald,Allen-Tildesley,ED75}. They are the analogs
of Einstein's formula for the self diffusion $D_s$, and are also
expressed in equilibrium time correlation functions, e.g.
\ba \Label{EH-TC}
D_s & = &  \lim_{t \to \infty} (1/2t)\av{(x(t)-x(0))^2}_0
\nn \eta &= &  \lim_{t \to \infty} (\beta/2
Vt)\av{(M(t)-M(0))^2}_0 ,
\ea
where $ \eta$ is the shear viscosity with $M(t)= \sum_i mv_{iy}(t)
x_i(t)$. In fact these forms give a more compact and more robust
representation of transport coefficient than the Green-Kubo
formulas \Ref{11} or \Ref{15}. This is the case because the former
can be expressed by only specifying the momentum and energy of a
particle, and do not require the explicit introduction of forces.
Consequently, the Helfand formulas apply to systems with
conservative, impulsive, dissipative and stochastic forces, and
either type of Green-Kubo formulas \Ref{11} or \Ref{15} can be
derived for the relevant cases. An added advantage of the
Einstein-Helfand formulas is that they are evidently positive.

So there exist two independent routes for generalizing Green-Kubo
formulas to non-conservative forces. The first one is to start
from the Einstein-Helfand formulas, and transform them into
\Ref{15}. This route has been followed in Ref.\cite{DE04}. The
second one is to derive the generalizations from the fundamental
Liouville equation, or its mesoscopic analog, the Fokker-Planck
equation. The latter route is followed in the present paper.

The plan of the paper is as follows: in Sec.II we obtain a
fundamental commutation relation between the $N-$particle
equilibrium distribution function $\rho_0$ and the infinitesimal
generator $\calL $ (Liouville or Fokker-Planck operator), that is
required in the derivation of the Green-Kubo formulas. In Sec.III
the generic form \Ref{15} of these formulas is derived using the
Zwanzig-Mori projection operator method in a form that is
applicable to systems with conservative forces (classical fluids,
lattice gas cellular automata, MPCD models), impulsive forces
(hard spheres), and stochastic and dissipative forces (Langevin
fluids). In Sec.IV we apply the new results to hard sphere fluids
with some technical details deferred to Appendix A. Further
comments and conclusions are presented in Sec.V.

\setcounter{equation}{0}
\section{Fundamental commutation relation}

We will analyze the similarities and differences among the
systems, mentioned in the introduction, all of which evolve to a
thermal equilibrium state described by the $N-$particle
equilibrium distribution function $\rho_0$. This enables us to
formulate a fundamental commutation relation for the infinitesimal
generator  $\calL$, that is required in the derivation of the
Green-Kubo formulas for systems whose equations of motion are not
time reversal invariant, or more precisely, whose pseudo-streaming
operators (Langevin fluids, elastic hard spheres) do not generate
time reversal invariant dynamics.

Suppose we have constructed for the system of interest the
infinitesimal generator $\calL$ and the streaming operator
$\exp(t\calL)$, that generates the time evolution of dynamic
variables {inside} averages. To study {equilibrium}  time
correlation functions, we need two different types of transposes
of $\cal L$. The first transpose, $\calLtilde$, referred to as
{\em adjoint}, is defined with respect to an  inner product with a
weight equal to unity, and a second transpose, $\calLinvert$,
referred to as {\em conjugate}, is defined with a weight $\rho_0$,
i.e.,
\ba\label{aa}
\int d\Gamma A\calL B & = & \int d\Gamma B\calLtilde A \nonumber \\
\av{A\calL B}_0 &=&  \int d\Gamma \rho_0 A\calL B\nonumber \\
&\equiv & \int d\Gamma \rho_0 B\calLinvert A = \av{B\calLinvert A}_0.
\ea
The corresponding streaming operators, $A(t)=\exp(t\calL) A(0)$,
and $A(-t)= \exp(t{\calL}^\epsilon)A(0)$ with $t>0$, generate the
{\it forward} and {\it backward} or time reversed evolution inside
averages in the sense that equilibrium time correlation functions
are {\it stationary}, i.e. depend only on time differences,
\ba\label{bb}
\av{A(0)B(t)}_0 &=& \av{Ae^{t\calL}B} \nonumber \\
= \av{Be^{t\tilde{\calL}} A}_0 &=& \av{A(-t) B(0)}_0.
\ea
Having obtained  a conjugate generator $\calL^\epsilon$, we also
introduce a conjugate current, $J_\epsilon =\lim_{k\to 0}
j_{\epsilon \bk}^*$, through the relations $\calLinvert a_\bk^*=
-i\bk\cdot \bj_{\epsilon\bk}^*$. For conservative forces the
$\calL$-operators are first order differential operators. So, the
adjoint $\tilde{\calL}=-\calL$. The conjugate operator
$\calLinvert$ for conservative interactions follows directly from
the second line in \Ref{aa}, because $\calL$ and $\rho_0$ commute,
and the conjugate operator equals the adjoint, i.e. for time
reversal invariant dynamics,
\be\Label{12}
\calL^\epsilon = -\calL =\calLtilde.
\ee
Note that here the conjugate current satisfies $J_\epsilon=J$.

In systems, such as lattice gas cellular automata and MPCD models,
the dynamics involves stochastic variables, but does not contain
any dissipative interactions. The microscopic equations of motion
{\it inside averages} are time reversal invariant. Here the
statistical averages include averages over all realizations of the
stochastic variables. The {\it time reversal} transform of the
Liouville operator $\calL^\epsilon$ satisfies the relation,
$\calL^\epsilon=-\calL$, as is the case for conservative systems.
This transform is obtained by reversing the sign of all variables
odd under the time reversal transformation (e.g. velocities). Of
course even variables remain  unchanged (e.g. positions,
energies).

Next, we consider the Langevin fluids. They are described by
$N$-particle Langevin equations \cite{Risken,Gardiner}, and reach
thermal equilibrium. From these equations  the corresponding
Fokker-Planck equation for the $N$-particle probability
distribution $\rho(\Gamma,t)$ is constructed, which is the
fundamental starting point for our linear response theory. The
latter is written here \cite{footnote} as $\partial_t
\rho=\calLtilde\rho$. As we are interested in time correlation
functions, it is sufficient to define streaming operators
$e^{t\calL}$ for the time evolution of dynamical variables {\it
inside} averages. This may be done through the adjoint,
\be\Label{12a}
\langle A(t)\rangle =\int d\Gamma A e^{t\calLtilde} \rho(\Gamma,0)=
\int d\Gamma \rho(\Gamma, 0) e^{t\calL} A = \langle e^{t\calL}A\rangle.
\ee
In this formulation  averages over the realizations of stochastic
forces are already accounted for in the Fokker-Planck equation.
So, $\calL$ is the adjoint of the Fokker-Planck operator
$\calLtilde$. Because of the presence of dissipative forces, the
Fokker-Planck equation is not time-reversal invariant.
Consequently, $\calLinvert\neq -\calL$ (no odd parity in the
velocities). The Fokker-Planck operator $\tilde{\calL}$ and its
adjoint ${\calL}$ contain first and second order derivatives, and
also here the adjoint can be simply obtained.

These Langevin equations have been constructed such that the
detailed balance condition holds by imposing that the Gibbs'
distribution is a steady state solution of the Fokker-Planck
equation, i.e. $\partial_t \rho_0 =\calLtilde\rho_0=0$. Then
dissipative and random forces are related by the
fluctuation-dissipation theorem. This implies that the
Fokker-Planck equation has an $H$-theorem or Lyapunov functional,
and that  $\rho_0 $ is the unique steady state solution.

Next consider the construction of the conjugate operator
$\calLinvert$, where $\rho_0$ and $\calL$ in \Ref{aa} do not commute. So
we introduce an operator $\calLbar$ through the commutation relation
\be\Label{13}
\rho_0 \calL =\calLbar\rho_0 .
\ee
Having obtained $\calLbar$, we insert this relation in line 2 of
\Ref{aa}, and obtain the conjugate generator as the adjoint of $\calLbar$,
i.e.
\be\Label{14}
\calLinvert =\widetilde {\calLbar}.
\ee
For the Fokker-Planck equations, satisfying the detailed balance
criteria, the commutation relation \Ref{13} has been derived in
the literature \cite{Risken,Gardiner} with a  general prescription
for $\calLbar$. Only in the case of conservative interactions,
where the equations of motion are time reversal invariant, the
equalities $\calLbar=\calL$ and $J_\epsilon=J$ hold. For
non-conservative interactions one has $\calLbar\neq \calL$ and
$J_\epsilon\neq J$.

As we will see in the next section, the replacement of \Ref{12}
for a time evolution {\em with} time reversal invariance, by the
commutation relation \Ref{13} for a time evolution {\em without},
has severe effects on the explicit form of the Green-Kubo formulas
for the Langevin fluids, as shown in \Ref{15}. There are two
conspicuous differences with respect to the standard Green-Kubo
formulas \Ref{11}. The fluxes $J_\epsilon$ and $J$ inside the
equilibrium time correlation function are necessarily {\it
unequal}, as $\calL_\epsilon \neq - \tilde{\calL}$, and there
appears an {\it additional} instantaneous transport coefficient
$\mu_\infty$, that solely depends on equilibrium properties.

Finally we discuss fluids of elastic hard spheres with mass $m$
and diameter $\sigma$. In the kinetic theory of hard sphere fluids
the standard notation \cite{EDHvL,vBE-JSP,dSEC-JSP,DE04} is:
$\calL = L_+$, and $\calL^\epsilon = -L_-$. Here the
pseudo-streaming operator $\exp[t\calL] = \exp[t L_+]$ with $t>0$
generates forward trajectories, and $\exp[t\calL^\epsilon] =
\exp[-t L_-]$ with $t>0$ backward or time reversed trajectories.
Furthermore, it has been shown \cite{EDHvL} that the correct hard
sphere dynamics inside an equilibrium average is generated by the
combination $ \rho_0 A(t) =\rho_0 e^{tL_+} A(0)$. Here the
presence of the factor $\rho_0$ guarantees that a zero weight is
given to the {\it unphysical} overlapping initial configurations,
for which the pseudo-streaming operators generate trajectories as
well. These trajectories are not time reversal invariant.

{From} the explicit form of the pseudo-Liouville operator $\calL$,
as briefly reviewed in Appendix A1, it is clear that $\calL$ is
{\it not} odd in the velocity variables, i.e. $\calL^\epsilon \neq
- \calL$, and consequently the trajectory, generated by the
streaming operator is not invariant under the time reversal
transformation. A typical example illustrating the lack of time
reversal invariance of the hard sphere trajectories, generated by
the streaming operators, is a trajectory in $\Gamma-$space
starting from an unphysical overlapping initial configuration.
Here the overlapping pairs {\it separate} within a finite time.
This forward trajectory is not retraced when all velocities are
reversed, because the pairs, when meeting again, experience a
standard hard sphere collision, in which the backward trajectory
is deflected from the forward one.

In the original literature \cite{EDHvL} it has also been shown that
the hard sphere generators obey for $t>0$ the {\em commutation}
relation,
\be\Label{16}
\rho_0e^{t\calL} =e^{t\calLbar}\rho_0,
\ee
which implies that the time correlation functions are stationary,
i.e. depend only a the time difference. In \Ref{A5} of Appendix A
the derivation of \Ref{16} is briefly recalled. In terms of time
correlation functions for dynamic variables $A$ and $B$ the
stationarity relation \Ref{bb} also applies, i.e.
\begin{equation}\Label{24a}
\av{A e^{t \calL}B}_0= \int d\Gamma A e^{t\calLbar}\rho_0  B =
\av{B e^{t \calLinvert} A}_0,
\end{equation}
where the transpose of $\bar{\calL}$  is $\calL^\epsilon$ on
account of \Ref{14}. Finally we observe that these relations also
include conservative systems as a special case, in which
$\bar{\calL} = \calL$ on account of \Ref{14} and \Ref{12}.
Consequently Eq.\Ref{13} reduces to the commutation relation,
$\rho_0\calL\cdots = \calL \rho_0\cdots $, implying that $\rho_0$
is a stationary solution of the Liouville equation,
$\partial_t\rho=\calLtilde\rho=-\calL\rho$. However,
$\calL\rho_0\neq 0$ for the Fokker-Planck equation, where
$\calLtilde\rho_0=0$ in the stationary case.

In {\it summary}, the streaming operators for Langevin and hard
sphere fluids both lack time reversal invariance, and both evolve
towards a thermal equilibrium state, and their time correlation
functions are stationary. These fundamental properties guarantee
both the commutation relation
\Ref{13} or \Ref{16}, as well as an identical structure of the
Green-Kubo formulas for such vastly different systems as hard
sphere fluids and Langevin fluids.

\setcounter{equation}{0}
\section{Zwanzig-Mori projection operator method}

The goal of this section is to derive for the fluid models,
described above, the hydrodynamic equations and Navier-Stokes
transport coefficients by means of the Zwanzig-Mori projection
operator method \cite{Hansen-McDonald,ED75}. This will be done by
deriving macroscopic equations for the averages of conserved
microscopic densities,  or their linear combinations, collectively
denoted by $a^i(\br)$,
\begin{equation} \Label{hydro-mode}
a^i(\br) = \{n(\br),g_\alpha(\br),e(\br)\},
\end{equation}
or by a subset of these densities. Here $n(\br)=\sum_i\delta
(\br-\br_i)$ is the number density, $\bg(\br)=\sum_i
m\bv_i\delta(\br-\br_i)$ the momentum density, and $e(\br) =\sum_i
e_i\delta(\br-\br_i)$ the energy density, where the single particle
energy $e_i=\half mv_i^2+\half\sum_{j(\neq i)}V(\br_{ij})$ or the
energy of an internal state.   In fact this program has been
carried out in Ref.\cite{ED75} for classical fluids with
conservative interactions. We mostly follow the notation and
definitions used in that article. Here we apply the Zwanzig-Mori
method to the generators of all systems, discussed in the previous
sections.

The general equation of motion for the Fourier-Laplace
transform of
$a^i(\br,t)$ is,
\begin{equation}\Label{29}
(z-\calL)a_{\bk z}=a_{\bk} ,
\end{equation}
and Fourier modes are defined as
\begin{eqnarray}\Label{210}
a_\bk & = & \int d\br e^{-i\bk\cdot\br} (a(\br)-
\langle a(\br)\rangle_0)\nonumber\\
a_{\bk z} & = & \int_0^\infty dt e^{-zt} a_\bk (t),
\end{eqnarray}
where $a^i_\bk$ is a linear combination of conserved densities,
which is for small $k$ orthogonal with respect to the {\it inner
product},
\begin{equation}\Label{211}
\langle a^i_\bk|a^j_\bk\rangle \equiv
V^{-1} \langle a^{i*}_\bk a^j_\bk\rangle_0 =\delta_{ij} +{\cal O}(k^2),
\end{equation}
and the asterisk denotes {\it complex conjugation}. This enables us
to define a projection operator $\calP =1-\calP_\perp$, as
\begin{equation}\Label{212}
\calP b_\bk = \sum_{\bq,j} a^j_\bq \langle a^j_\bq|b_\bk\rangle
= \sum_j a^j_\bk \langle a^j_\bk|b_\bk\rangle.
\end{equation}
Here $\langle a^j_\bq|b_\bk\rangle = \delta_{\bq,\bk}\langle
a^j_\bk|b_\bk\rangle $ because of translational invariance.
Applications of the Zwanzig-Mori method to ~(\ref{29}) yields for
the projected part $\calP a_{\bk z}$,
\begin{equation}\Label{213}
[z-\calP\calL\calP -\calP\calL\calP_\perp(z-
\calP_\perp\calL\calP_\perp)^{-1}\calP_\perp\calL\calP]
\calP a^m_{\bk z}= a^m_\bk,
\end{equation}
or in terms of hydrodynamic propagators, $G_{ij}(\bk,z)=\langle
a^i_\bk|a^j_{\bk z}\rangle$, we obtain,
\begin{equation}\Label{214}
\sum_j\left[z\delta_{ij} +ik\Omega_{ij}(\bk)
+k^2 U_{ij}(\bk,z)\right]G_{jm}(\bk,z)=\delta_{im} .
\end{equation}
The hydrodynamic matrix, $H_{ij} = ik\Omega_{ij} + k^2 U_{ij}$,
consists of an equilibrium average, $\Omega_{ij}$, which is the
Euler matrix, and a time correlation  function, $U_{ij}$, which is
the Navier-Stokes transport matrix. Its real part, $\Re
H_{ij}(\bk,z)$, is of ${\cal O}(k^2)$, and represents the
dissipative part, and contains the Navier-Stokes transport
coefficients of shear viscosity $\eta$, bulk viscosity $\zeta$ and
heat conductivity $\lambda$. For conservative forces $\Im H_{ij} =
k \Omega_{ij}$ and $ \Re H_{ij} = k^2 U_{ij}$. For
non-conservative forces $\Omega_{ij}$ contains real and imaginary
parts, as we shall see. By diagonalizing the hydrodynamic matrix
for small $k$ to $\calO (k^2)$ one finds the hydrodynamic
eigenvalues in the form, $z_\lambda \simeq -ik c_\lambda -k^2
D_\lambda$, and the hydrodynamic modes, $a^i_\bk$, as  special
linear combinations of conserved densities. They will be made
explicit below. Here the imaginary part of $z_\lambda$ contains
the propagation speed $c_\lambda$ of that mode (which may be
vanishing), and the real part contains the diffusivity or dampings
coefficient $D_\lambda$ of the mode. We cite the results of
equations (44) and (45), obtained in Ref.~\cite{ED75},
\begin{eqnarray}\Label{215}
D_\eta &=& {\eta}/{\rho} =\calL im \;
\Re \left\{ ik\Omega_{\eta\eta}(\bk)+
k^2U_{\eta\eta}(\bk,z)\right\}/k^2 \nonumber \\
&=&  -\calL im \; \Re \left\{ \langle a_\bk^{\eta_i} |\calL
a_\bk^{\eta_i}\rangle + \langle \calL^\epsilon a_\bk^{\eta_i} |
\calP_\perp \hat {\cal G}_z \calP_\perp \calL
a_\bk^{\eta_i}\rangle \right\} /k^2  \nonumber \\
D_l  &=& -\calL im \; \Re \left\{ \langle a_\bk^{l} |\calL
a_\bk^{l}\rangle + \langle \calL^\epsilon a_\bk^{l} | \calP_\perp
\hat {\cal G}_z \calP_\perp \calL
a_\bk^{l}\rangle \right\}  /k^2  \nonumber \\
D_T ={\lambda}/({n{\cal C}_p})&=&   -\calL im  \; \Re \left\{
\langle a_\bk^{T} |\calL a_\bk^{T}\rangle + \langle \calL^\epsilon
a_\bk^{T} | \calP_\perp \hat {\cal  G}_z \calP_\perp \calL
a_\bk^{T}\rangle
\right\}/k^2.
\end{eqnarray}
Here the symbols $D_\eta,\,D_l$ and $D_T$ represent respectively
the transverse, longitudinal and thermal diffusivities, and
$\rho=mn$ is the mass density and ${\cal C}_p$ is the specific heat
per particle at constant pressure. Furthermore, $\calL im$
represents the double limit, $\lim_{z\to 0}\lim_{k\to 0}$. In the
equation above $\hat {\cal G}_z$ is the projected resolvent
operator,
\begin{equation}\Label{215b}
\hat\calG_z= \calP_\perp
(z-\calP_\perp\calL\calP_\perp)^{-1}\calP_\perp ,
\end{equation}
which reduces in the small-$k-$limit to the standard resolvent, $
\calP_\bot (z-\calL)^{-1}\calP_\bot$, when acting on {\it
projected} dynamic variables \cite{ED75}. The transformation of
the matrix element $U_{ij}(\bk,z)$ from the representation
$\av{a_\bk^i|\calL\calP_\perp \cdots }$ in \Ref{213} to the
representation $\av{\calLinvert a_\bk^i|\calP_\perp\cdots}$ in
\Ref{215} follows directly from the commutation relation \Ref{13},
valid for Langevin- and hard sphere fluids. We also note that the
relation applies to systems with dynamics that is {\em time
reversal invariant}, i.e. where $\calL^\epsilon = -\calL$.

In {\it summary}, the equations as listed in
\Ref{215} are valid for all models discussed in this paper.
Through this formulation we have extended the linear response
theory for conservative systems in Ref.\cite{ED75} to more complex
fluids like Langevin and hard sphere fluids.

In order to apply these formulas to different systems the
definitions and some properties of the quantities appearing in
\Ref{215} will be required. Due to the isotropy of simple
fluids in thermal equilibrium the $(\eta_i, \eta_j)-$matrix
elements  between shear modes  $\{a^{\eta_i}_\bk\}$ are diagonal
and {\it independent} of the transverse directions $(i=1,2,\cdots,
d-1)$.
 We further recall that
the viscosity in isotropic  simple fluids is a fourth rank
isotropic tensor, determined by two independent scalars, i.e.
\begin{equation}\Label{215c}
\eta_{\alpha\beta\gamma\delta}=
\eta\{\delta_{\alpha\gamma}\delta_{\beta\delta}+
\delta_{\alpha\delta}\delta_{\beta\gamma}-
({2}/{d})\delta_{\alpha\beta }\delta_{\gamma\delta}\} +\zeta
\delta_{\alpha\beta }\delta_{\gamma\delta}.
\end{equation}
They are  the shear- and bulk viscosity,  $\eta$ and $\zeta$
respectively, or alternatively the transverse and longitudinal
diffusivity, $D_\eta$ and $D_l$, determined by,
\begin{eqnarray}\Label{215d}
\eta_{xyxy} & = & \rho D_\eta =\eta \nonumber \\
\eta_{xxxx} & = & \rho D_l = 2\eta (1-d^{-1}) + \zeta.
\end{eqnarray}
For fluids with conserved densities $n(\br),g_\alpha(\br)$ and
$e(\br)$ --- i.e. classical fluids with conservative interactions,
and hard sphere fluids with impulsive interactions---, the explicit
expressions for the hydrodynamic modes are \cite{Resibois,ED75},
\begin{eqnarray}\Label{217}
a_\bk^{\eta_i} & = & \sqrt{\frac{\beta}{\rho}}\,
\bg_\bk\cdot \hat{\bk}_\perp^i \
(i=1,2,\cdots,d-1)\nonumber\\
a_\bk^\sigma & = & \sqrt{\frac{\beta}{2\rho}}
\left\{\frac{1}{c} p_\bk +\sigma g^l_\bk\right\}\
(\sigma=\pm 1)\nonumber\\
a_\bk^T & = & \sqrt{\frac{\beta}{nT{\cal C}_p}}\left\{e_\bk
-hn_\bk\right\},
\end{eqnarray}
where $c$ is the adiabatic speed of sound with $c^2 = (\partial
p/\partial \rho)_s$, and $p_\bk$ is the thermodynamic pressure
fluctuation,
\begin{equation}\Label{218}
p_\bk=\left(\frac{\partial p}{\partial n}\right)_e n_\bk +
\left(\frac{\partial p}{\partial e}\right)_n e_\bk.
\end{equation}
Furthermore $\{\hat\bk,\hat\bk_\perp^i\}\, (i=1,2,\dots,d-1)$ is a
set of $d$ ortho-normal unit vectors, where the longitudinal
direction $\hat\bk$ is taken parallel to the $x$-axis for
convenience. Moreover, $g^l_\bk = \hat \bk \cdot \bg_\bk$ is the
longitudinal component of $\bg_\bk$. Here $a_\bk^{\eta_i}$ are the
shear or transverse velocity modes, $a_\bk^\sigma\, (\sigma=\pm)$
the sound modes, and $a_\bk^T$ the heat mode.

It remains to work out the expressions (\ref{215}) for the
different fluid models, using the appropriate hydrodynamic modes
$a_\bk^i$ in  (\ref{217}), in combination with the explicit forms
in the pseudo-Liouville operator for these models.

Equations \Ref{215} are the main new results of the general part
of this paper. We note the latent presence of the structure of
\Ref{15}, where the instantaneous transport coefficient
$\mu_\infty \propto  \Re\av{a_\bk|\calL a_\bk}$, and the time
correlation functions with the different currents, $J_\epsilon$
and $J$, are contained in the  matrix $U(\bk,z)$. Inserting the
currents, defined for small $k$ through,
\ba \Label{219}
\calL a_\bk &=& -ik J + \calO (k^2) \nn
\calL^\epsilon a_\bk ^*&=& - ik J_{\epsilon}+ \calO (k^2),
\ea
into \Ref{215} reduces this equation to the generic structure
\Ref{15}, and we recall that the inner product in \Ref{215}
involves complex conjugation. Consider first the case of {\it
conservative} forces. There the dynamics is time reversal
invariant, $ \calL^\epsilon  = -\calL $, i.e. $\calL$ is odd in
the velocity variables, and the contributions of the real part of
the $\Omega-$ matrix necessarily vanishes, i.e. $\mu_\infty =0$.
Moreover, both currents, as defined in \Ref{219}, are equal,
$J_{\epsilon} =  J$.

In case the forces are {\it not} conservative, i.e.
$\calL^\epsilon \neq -\calL $, then $\mu_\infty \propto \Re k^{-2}
\av{a_\bk|\calL a_\bk}$ is necessarily {\it non-vanishing}, and
approaches a finite value in the small $k-$limit. Moreover, it
follows from general considerations that $J_{\epsilon}$ and $J$
are {\it unequal}, both for {\it impulsive}  forces (hard sphere
fluids), and for {\it dissipative} forces (Langevin fluids). The
explicit expressions will be discussed in \cite{ME-Langevin}.

In hindsight the instantaneous transport coefficients $\mu_\infty$
can be identified independently of the present analysis. In a
kinetic theory description, $\mu_\infty$ comes from the
irreversible part of the average of the microscopic or mesoscopic
$N$-particle current $\av{J}_{loc}$ in a state of {\em local
equilibrium}, i.e. the part of $\av{J}_{loc}$ that contributes to
the {\em irreversible entropy production}. In the context of
complex fluids this is the part that contributes to the
Navier-Stokes transport matrix in the hydrodynamic equations. For
instance, in the momentum current or stress tensor one finds both
in the hard sphere fluid \cite{Resibois,Chapman-Cowling}, as well
as in Langevin fluids \cite{Marsh}, $\av{J_{xy}}_{loc}=\rho u_x
u_y -\eta_\infty \nabla_x u_y$. Here the first term, $\rho
u_xu_y$, is the Euler or reversible part of the current, and the
second term is the irreversible part, yielding a contribution
$\eta_\infty$ to the Navier-Stokes transport coefficients. For the
hard sphere fluid the statement above will be explicitly verified
in Sect. IV. For the Langevin fluids and solids it will be
verified in Ref. \cite{ME-Langevin}. As the concept of local
equilibrium state is intuitively clear, one can in principle also
estimate $\mu_\infty$ for more complex fluids.

Perhaps the most intuitive understanding of $\mu_\infty$ has been
given by Hoogerbrugge and Koelman \cite{Koelman} when first
introducing the dissipative particle dynamics (DPD)-model. The
estimate of the shear viscosity of the DPD-fluid $\eta\simeq
\eta_\infty$ by making a continuum (macroscopic) approximation of
the Langevin equation of motion.
We finally note that the instantaneous transport coefficient is
also one of the main contributions to the transport coefficients
of hard sphere- and Langevin fluids at liquid densities.

The appearance of two different currents $J$ and $J_\epsilon$ in
the Green-Kubo formulas for the viscosity and friction coefficient
in hard sphere fluids have been derived before in Refs.
\cite{DE04,Pias}. For an gas of inelastic hard spheres similar
results have been obtained in Ref. \cite{Dufty-Baskaran05}. For a
general class of Fokker-Planck equations, satisfying detailed
balance, local macroscopic evolution equations have been derived
in Ref.\cite{pep-Vasquez}. In doing so the expressions, derived
for the transport coefficients, are also given in the form of
Green-Kubo formulas with two different currents $J$ and
$J_\epsilon$. However, the relation between $J_\epsilon$ and $J$
in that case is more complicated than in the cases discussed in
the present paper.

This ends the general part of the paper. The remainder deals with
the most important application of the present theory, the hard
sphere fluid.

\setcounter{equation}{0}
\section{Classical and hard sphere fluids}
\subsection{Classical conservative fluids}

In the application part of this paper we first list for later
reference the Green-Kubo formulas in conservative classical fluids,
and next we work out the details of the Green-Kubo formulas for
hard sphere fluids. For classical fluids with conservative
interactions the explicit forms have been derived in
\cite{ED75}. Here the Liouville operator is odd in the velocities
because of time reversal invariance. Consequently the diagonal
elements $\Omega_{\eta\eta}$ and $\Omega_{TT}$ of the Euler matrix
in (\ref{215}) are vanishing. Computation of the elements of the
Navier-Stokes matrix yields  for the shear viscosity $\eta$,
longitudinal viscosity $D_l$ and heat conductivity $\lambda$
\cite{ED75},
\ba\Label{31}
\eta &=& ({\beta}/{V}) \int_0^\infty\! dt
\langle \delta S_{xy}(0) \delta S_{xy}(t)\rangle_0
\equiv \int^\infty_0 dt C_\eta (t) \nn
\rho D_l &=& ({\beta}/{V})\int_0^\infty\! dt
\langle \delta S_{xx}(0) \delta S_{xx}(t)\rangle_0
\equiv \int^\infty_0 dt C_l (t) \nn
\lambda &=& ({\beta}/{TV})\int_0^\infty dt
\langle Q_x(0) Q_x(t)\rangle_0 \equiv \int^\infty_0 dt C_T (t)
\ea
with the subtracted microscopic fluxes,
\ba \Label{32}
\delta S_{xy}&=& \calP_\bot J_{xy} =J_{xy} =
\sum_imv_{ix}v_{iy} +\sum_{i<j} r_{ij,x}F_{ij,y}
\nn \delta S_{xx}&=& \calP_\bot (J_{xx} -\av{J_{xx}}_0)
\nn & =& J_{xx} -pV -\left(\frac{\partial p}{\partial e}\right)_n
(H -\av{H}_0) -\left(\frac{\partial p}{\partial n}\right)_e
(N-\av{N}_0)
\nn  Q_x & = & \calP_\perp J_x =  J_x-P_x \av{P_x J_x}_0 /\langle
P_x P_x\rangle_0 = J_x-\frac{h}{m}P_x
\nn J_x & = & \sum_i e_i v_{ix} +
\sum_{i<j}(\bv_i+\bv_j)\cdot \bF_{ij}r_{ij,x} .
\ea
Here  $J_{x\alpha}$ is the microscopic $N$-particle momentum flux
or stress tensor, where $F_{ij,\alpha} = -\partial V(r_{ij})
/\partial r_{ij,\alpha}$ with $\alpha=x,y,\dots,d$ is the
conservative inter-particle force and $V(r_{ij})$ the spherically
symmetric pair potential, $\av{J_{xx}}_0 = pV$ is the hard sphere
pressure in equilibrium, $h =(e+p)/n$ is the enthalpy per particle,
and $P_\alpha =\sum_i m v_{i\alpha}$  the total momentum. Moreover,
$J_\alpha$ is the total energy flux. The quantity $P_\alpha h/m$,
which is the component of $J_ \alpha$ parallel to the conserved
quantity $P_\alpha$, is subtracted out in $Q_x$ \cite{ED75}.

\subsection{Hard spheres: Instantaneous transport coefficients}

This section deals with the $\Omega$ matrix, $\av{a_\bk^i|\calL
a_\bk^i}$, in which the first major difference in the Green-Kubo
formulas shows up between systems with time reversal invariance,
i.e. $\;\calLinvert = -\calL$, and those without, i.e. $\calLinvert
\neq -\calL$. So, in hard sphere systems the Liouville
operators necessarily contain even parts in the velocities---apart
from possible odd parts. The matrix elements of $\calL$ always have
real parts of ${\cal O}(k^2)$, that contribute to the irreversible
or Navier-Stokes part of the hydrodynamic equations. These terms
are referred to as the {\it instantaneous} or high frequency parts
of the transport coefficients. The appearance of such terms is a
very robust feature of all systems with streaming operator that
lack time reversal invariance.   We also recall that  the streaming
operator $\exp( t \calL)$ in \Ref{215} have to be defined {\it for
all} points in phase space, including the physically inaccessible
overlapping ones.

For hard sphere fluids the instantaneous contributions, which are
averages in thermal equilibrium, are calculated in Appendix A2, and
yield results, {\it exact} for a $d$-dimensional hard sphere fluids
at general densities,
\begin{eqnarray}\Label{33}
\eta_\infty & = &\varpi d/(d+2)  \nn
\lambda_\infty & = &\varpi dk_B/(2m) \equiv{\cal C}_V \varpi \nn
\rho D_{l \infty} &=& 3 \varpi d/(d+2)  \nn
\zeta_\infty  &=& \varpi.
\end{eqnarray}
For completeness we have also listed the instantaneous contribution
to the longitudinal diffusivity and the bulk viscosity, which have
been obtained similarly with the help of \Ref{215d}. In the
equations above we have introduced
\begin{eqnarray}\Label{34}
\varpi & = & \rho\sigma^2/ d^2t_E \nn
t_E & = & \sigma \sqrt{\pi\beta m} /2dbn\chi,
\end{eqnarray}
where the definition of $\varpi$ has been chosen such that it
coincides for $d=3$ with the notation used in Ref.
\cite{Chapman-Cowling},  and $t_E$ is the Enskog mean free time,
$\chi=g_0(r=\sigma^+)$ is the value of the radial distribution at
contact, $b=(1/2d) \Omega_d \sigma^d$ is the excluded volume which
equals half the volume of an action sphere with radius  $\sigma$,
and $\Omega_d=2\pi^{d/2}/\Gamma(d/2)$ is the surface area of a
$d$-dimensional unit sphere.  Alternatively these parameters may
be expressed in terms of the equation of state for hard spheres
through $\beta p = n(1+bn\chi)$.

\subsection{Hard spheres: time correlation functions}

The Navier-Stokes matrix $\lim_{\bk\to 0}U(\bk,z)= U(z)$ gives in
the long wave length limit the frequency dependent ($ z= i\omega
$) transport coefficients (memory effects). In the previous
subsection we have been dealing with the contributions
$\eta_\infty, \zeta_\infty, \lambda_\infty$ at infinite frequency.
We introduce $\delta C_a (t)$ to denote the inverse Laplace
transform of the Navier-Stokes part of the hydrodynamic matrix, $
\rho U_{aa}(z) \equiv \delta \tilde{C}_a (z)$ with $a =\{
\eta,l,T\}$. Using the explicit forms (\ref{217}) for the
hydrodynamic modes we obtain from (\ref{215}) the Laplace
transforms of the time correlation functions,
\begin{eqnarray}\Label{35}
\rho U_{\eta\eta}(z) \equiv \delta \tilde C_\eta(z) &=&
- \lim_{k\to0} \frac{\beta}{k^2V}
\langle (\calL^\epsilon g^*_{\bk y}) \calP_\perp \hat {\cal G}_z
\calP_\perp \calL g_{\bk y}\rangle _0
\nn  \rho U_{ll}(z) \equiv \delta \tilde C_l(z) &=&
- \lim_{k\to0} \frac{\beta}{k^2V}
\langle (\calL^\epsilon g^*_{\bk x}) \calP_\perp \hat {\cal G}_z
\calP_\perp \calL g_{\bk x}\rangle _0
\nn n{\cal C}_p U_{TT} (z) \equiv \delta \tilde C_T(z) &=& -
\lim_{k\to 0} \frac{\beta}{k^2TV}
\langle (\calL^\epsilon e^*_{\bk}) \calP_\perp
\hat {\cal G}_z \calP_\perp \calL e_{\bk}\rangle _0.
\end{eqnarray}
The asterisk denotes complex conjugation. From here on only the
standard notation of hard sphere kinetic theory
\cite{EDHvL,vBE-JSP,dSEC-JSP,DE04}  will be used with
$\calL=L_+$ and $\calL^\epsilon = -L_-$. To compute the explicit
form of the currents we use the relation,
\be\Label{36a}
 \bk b_\bk = \bk \delta b_\bk =
  \bk (b_\bk -\delta_{\bk,\mbox{\scriptsize\boldmath $0$}} \av{b_{\mbox{\scriptsize\boldmath $0$}}}_0)
 = \bk \delta b_{\mbox{\scriptsize\boldmath $0$}} +\calO(k^2),
\ee
where  the fluctuation $\delta b_\bk = b_\bk -\av{b_\bk}_0 $. Here
we have used the relation  $\av{b_\bk}_0 =
\delta_{\bk,\mbox{\scriptsize\boldmath $0$}}\av{b_{\mbox{\scriptsize\boldmath $0$}}}_0 $,
which is a consequence of translational invariance. The
replacement of $ b_\bk$ by $\delta b_\bk$ in
\Ref{36a} is convenient for taking the limit $\bk \to {\bf 0}$. It is
further convenient to choose $\bk$ parallel to the $x$-axis.
Consider first the longitudinal $(\alpha = x)$ and transversal
$(\alpha = y)$ momentum current. Then we obtain  with the help of
\Ref{36a} for small $k$,
\begin{eqnarray}\Label{36}
\calL g_{\bk \alpha} & = & L_+g_{\bk \alpha} = -ik \delta J_{+x\alpha}
 = -ik (\delta J_{x\alpha}^k
+\delta J_{+x\alpha}^v) \nn
\calL^\epsilon g_{\bk \alpha}^* & = &  -L_- g_{\bk \alpha}^*
= -ik \delta J_{-x\alpha} = -ik (\delta J_{x\alpha}^k +\delta
J_{-x\alpha}^v) .
\end{eqnarray}
By inserting the expressions for the hard sphere generators in
\Ref{A1}, and using the properties \Ref{A2}-\Ref{A7} it is
straightforward to obtain the kinetic part ($k$) and collisional
transfer part ($v$) of the momentum currents,
\begin{eqnarray}\Label{37}
J_{x\alpha}^k & = & \sum_i mv_{ix} v_{i\alpha} \nonumber \\
J_{\pm x\alpha}^v & = & \pm \sum_{i<j} T_{\pm}(ij) m
(v_{i\alpha} r_{ix} + v_{j\alpha}r_{jx}) \nonumber \\
   & = & \sum_{i<j} m\sigma^d \int^{(\mp)}
   d\hat\sigma (\bv_{ij}\cdot \hat\bsigma)^2
   \hat\sigma_x\hat\sigma_\alpha
\delta(\br_{ij} -\sigma\hat\bsigma).
\end{eqnarray}
 The averages of the fluxes are only {\it non-vanishing }
for $\alpha = x$, i.e.
\ba \Label{37a}
\av{J^k_{xx}}_0 & =& V n /\beta, \quad \mbox{and} \quad
\av{J^v_{\pm xx}}_0  = Vn^2 b \chi /\beta
\nn
\av{J_{\pm xx}}_0 & = & Vn(1+b n \chi)/\beta = Vp
\ea
in complete agreement with the result \Ref{32} for smooth
potentials. In the transition from the second to the third line in
\Ref{37} one may use \Ref{A9} and \Ref{A2} and express the
variables $\{\bv_i,\br_i,\bv_j,\br_j\}$ in center of mass
coordinates $\{\bR_{ij},\bG_{ij}\}$, and relative ones
$\{\br_{ij},\bv_{ij}\}$, and apply \Ref{A7} for conserved
variables. Consider first the transverse stress fluctuation with
$\alpha = y$. Here we note that $\calP_\perp J_{\pm xy}=J_{\pm xy}$
in \Ref{35}, as the $(xy)$-component of the microscopic stress
tensor has no subtracted parts. The momentum fluxes,  $J_{+xy}^v$
and $J_{-xy}^v$, differ only from one another in the constraints on
the integral over the $d$-dimensional solid angle $d\hat\sigma$,
denoted by the superscript ($-$) or ($+$). Here ($\mp$) requires
$(\mp)
\bv_{ij}\cdot \hat\sigma >0$, and implies that the
$\hat\sigma$-integration is restricted to the pre-collision
hemisphere ($-$), or to the post-collision hemisphere ($+$).

The time correlation function for the stresses for $t>0$ can now be
written as
\begin{equation}\Label{38}
\delta C_\eta(t) =\frac{\beta}{V} \av{J_{-xy}e^{tL_+} J_{+xy} }_0
= \frac{\beta}{V} \av{J_{+xy}e^{-tL_-} J_{-xy} }_0
=\sum_{a,b}\delta C^{ab}_\eta (t),
\end{equation}
where $a,b=\{k,v\}$. The second equality has been obtained by
renaming the dummy variables $\bv_i$ in $-\bv_i$ and using
\Ref{A1}. Also note that the cross-correlations are equal, $\delta
C^{kv}_\eta(t)= \delta C^{vk}_\eta(t)$, as follows directly from
the commutation relation
\Ref{24a} for hard sphere fluids, where $\calL =L_+$ and
$\calL^\epsilon = -L_-$, i.e.
\be \Label{HS-symm}
\av{J^k_{xy} e^{tL_+} J^v_{+xy}}_0 = \av{J^v_{+xy} e^{-tL_-} J^k_{xy}}_0
= \av{J^v_{-xy} e^{tL_+} J^k_{xy}}_0.
\ee
The last equality has again been obtained by renaming the dummy
integration variables $\bv_i$ to $-\bv_i$. The final Green-Kubo
formula for the shear viscosity in a hard sphere fluid follows then
from (\ref{215}), by taking the small-$z$ limit of the frequency
dependent $\delta \tilde{C}_\eta(z)$ and combining it with the
instantaneous viscosity in \Ref{33} to yield  the equivalent
representations of the generic structure \Ref{15},
\be \Label{eta-total}
\eta = \eta_\infty +  ({\beta}/{V})\int^\infty_0\!  dt\,
\av{J_{-xy} e^{tL_+}J_{+xy} }_0.
\ee
The hard sphere result for the shear viscosity as well as
(\ref{36})-(\ref{38}) have been derived already in
Ref.\cite{DE04}, by re-interpreting the Einstein-Helfand formulas
in terms of pseudo-Liouville propagators.

For the heat conductivity the analysis proceeds along similar
lines, and the energy flux is
calculated from the Fourier mode of the energy density $e_k$, in
\Ref{hydro-mode} with $e_i=\half m v_i^2$, i.e.
\begin{equation}\Label{39}
\calL e_\bk = -ik J_{+x} = -ik (J^k_x +J^v_{+x}),
\end{equation}
and a relation for $J_{-x}$ similar to the second line in
(\ref{36}). The subtracted flux is the same as in \Ref{32}
for conservative fluids, where the enthalpy per particle for hard
spheres is $h=h^k +h^v$ with $e=\half d n/\beta$  and $p =
(n/\beta)(1+bn\chi)$, and where we have used the relation $\av{P_x
J_{\pm x}} = Vn^2 b\chi/\beta^2 = h^v nV/\beta$. The kinetic and
collisional parts are then given by
\begin{eqnarray}\Label{310}
&& Q^k_x = {\calP}_\perp J_x^k  =  \sum_i\half [m v_i^2
-(d+2)/\beta] v_{ix}
\nn &&Q^v_{\pm x} = \calP_{\perp} J_{\pm x}^v = \pm \half
m\sum_{i<j} T_{\pm} (ij) (v_i^2 r_{ix}+ v_j^2 r_{jx}) -(h^v/m)P_x
\nn
 && =  \sum_{i<j} m \sigma^d \int^{(\mp)} d\hat\bsigma
 ({\bf v}_{ij}\cdot\hat\bsigma )^2
 (\bG_{ij}\cdot\hat\bsigma)\hat\sigma_x
\delta(\br_{ij}- \sigma\hat\bsigma) - (bn\chi/\beta m)P_x,
\end{eqnarray}
where $\bG_{ij}=\half (\bv_i+\bv_j)$. The time correlation function
of the subtracted heat flux is then,
\begin{equation}\Label{311}
\delta C_T(t) = \frac{\beta}{TV} \av{Q_{-x}e^{tL_+}Q_{+x} }_0
=\sum_{ab} \delta C^{ab}_T (t),
\end{equation}
and the cross-correlations $(kv)$ and $(vk)$ are again equal (see
\Ref{HS-symm}). The total heat conductivity in a hard sphere
fluid is then similarly given by the Green-Kubo formula,
\be \Label{312}
\lambda = \lambda_\infty +  ({\beta}/{TV})\int_0^\infty dt  \langle
Q_{-x} e^{tL_+} Q_{+x} \rangle_0 ,
\ee
which has again the generic structure \Ref{15}. Obviously, this
formula can also be derived from the Einstein-Helfand formulas,
following the method of \cite{DE04}.

For the longitudinal diffusivity $\rho D_{l}$ and the total bulk
viscosity $\zeta$ we obtain similarly with the help of \Ref{36a}
-\Ref{37a},
\ba \Label{313}
\rho D_l &=& \rho D_{l\infty}+  ({\beta}/{V})\int^\infty_0 dt \av{\delta
S_{-xx}e^{t L_+} \delta S_{+xx}}_0 \nn
\zeta &=& \zeta_\infty + ({\beta}/{V}) \int^\infty_0 dt \av{\delta
S_{-}e^{t L_+} \delta S_{+}}_0,
\ea
where the subtracted fluxes are,
\ba \Label{314}
\delta S_{\pm} &=&\frac{1}{d} \sum_\alpha \delta S_{\pm\alpha\alpha} \nn
\delta S_{\pm xx} &=&{\cal P}_\bot \delta J_{\pm xx} \nn
&=& \delta J_{+xx} - \left(\frac{\partial p}{\partial e}\right)_n
\delta H - \left(\frac{\partial p}{\partial n}\right)_e \delta N,
\ea
in complete agreement with the subtracted fluxes in \Ref{32}.
Before concluding this section we note that the instantaneous or
high frequency transport coefficients are also present in the
Enskog theory for dense hard sphere fluids, as the {\it local
equilibrium} averages of the {\it collisional transfer} parts in
\Ref{37} and \Ref{310} for the corresponding currents, i.e.
\ba \Label{315}
\av{J^v_{+xy}}_{loc} &=& -\eta_\infty \nabla_x u_y \nn
\av{J^v_{+x}}_{loc} &=&  -\lambda_\infty \nabla_x T \nn
\sum_\alpha \av{J^v_{+\alpha\alpha}}_{loc} &=& p -\zeta_\infty
\mbox{\boldmath $\nabla$} \cdot {\mbox{\boldmath $u$}},
\ea
where the bulk viscosity appears as a correction to the local
equilibrium hard sphere pressure $p$, and $\bf u$ is the local
flow field. As the collisional transfer terms are sums over pairs
of particles, the averages contain the pair distribution function
in local equilibrium, and involves integrations over $\{
\bv_1,\bv_2,\br_{12}\}$. They have been calculated in
Refs.\cite{Resibois,Chapman-Cowling,RET} for the special case of
$d=3$ as $\eta = (3/5)\varpi$, $\lambda = {\cal C}_ V\varpi$ and
$\zeta= \varpi$, and can for instance be found in Eqs.(16.14,8),
(16.42,3) and (16.41.7)) of Ref.\cite{Chapman-Cowling} in complete
agreement with the results derived here. The explicit expression
for $ \varpi $ in \Ref{34} is proportional to $n^2$. Consequently
at liquid densities the instantaneous transport coefficients are
major contributions, and account for about half the value of the
transport coefficient.

For comparison we quote the complete prediction of the Enskog
theory for the viscosities of a $d-$dimensional hard sphere fluid.
For the bulk viscosity one has $\zeta =\zeta_\infty = \varpi$, and
for the shear viscosity,
\be \Label{eta-Enskog}
\eta =\frac{1}{ \chi}\left( 1+ \frac{2}{d+2} bn\chi\right)^2 \eta_0
+ \frac{d}{d+2}\varpi.
\ee
Here $\eta_0$ is the low density Boltzmann value of the shear
viscosity \cite{Chapman-Cowling}. The expression above shows that
the Enskog theory gives for the shear viscosity other contributions
of order $n^2$ at high densities, and that the Enskog prediction
for the instantaneous viscosity $\eta_\infty = d\varpi/(d+2)$ is
exact for all densities.

\setcounter{equation}{0}
\section{Conclusions}

The main result of this paper is a generalization of the standard
Green-Kubo or time correlation expressions in \Ref{215} for
classical fluids with conservative interactions to hard sphere
fluids with impulsive interactions, and to mesoscopic Langevin
fluids with dissipative and stochastic interactions.

In the classes of fluids above the time dependence of dynamical
variables, $A(t)=\exp[t\calL]A(0)$, can be represented by
streaming operators or generators $\exp[t\calL]$ when used inside
ensemble averages. In conservative systems the dynamics generated
by the (pseudo-)streaming operators is time reversal invariant,
implying $\calLinvert=-\calL$ (see \Ref{12}). This property is
necessary to obtain Green-Kubo formulas of the standard form
\Ref{11}. In the remaining classes the dynamics generated by the
(pseudo-)streaming operator is not time reversal invariant, i.e.
  the backward streaming operator $\exp(t
\calL^\epsilon) \neq \exp( - t\calL)$ (with $t>0$).

The common criteria, obeyed by the models above are that the
systems approach a thermal equilibrium state,
$\rho_0\sim\exp[-\beta H]$, and that the corresponding correlation
functions $\av{A(t)B(t')}_0$ are stationary, i.e. only depend on
time differences. The criteria imply that the infinitesimal
generators for hard sphere and Langevin fluids obey the commutation
relation \Ref{13}, i.e. $\rho_0\calL=\calLbar\rho_0$. This includes
in fact conservative systems as the special case with
$\calLbar=\calL$.

There are two conspicuous differences between Green-Kubo formulas
in these three classes of fluids. In the classes without time
reversal invariant dynamics, i.e. $\calLinvert
\neq -\calL$, there is necessarily an instantaneous non-vanishing
contribution $\mu_\infty$, and in the classes with, i.e.
$\calLinvert = -\calL$, the generator is odd in the velocity
variables, and $\mu_\infty$ automatically vanishes (compare
\Ref{eta-total} and
\Ref{312} with
\Ref{31}). The next difference concerns the microscopic fluxes. In
the classes with time reversal invariance the Green-Kubo formula in
\Ref{11} is a time integral over an {\em auto}-correlation function
$\av{J(0)J(t)}_0$, containing the same currents. In the cases
without, the two currents in $\av{J_\epsilon e^{t\calL} J}_0$ are
unequal. Both types of differences apply to the hard sphere fluids,
as well as the Langevin fluids. In the former case our results
confirm and extend the results in \cite{DE04}.

A summary of the Green-Kubo formulas for transport coefficients in
Langevin fluids can be found in Ref.\cite{EB-EPL}, and a detailed
analysis will be given in Ref.\cite{ME-Langevin}, where the results
\Ref{215} will be applied to a Langevin-type solid showing heat
conduction, and a ditto fluid showing viscosity.

Finally, to understand the origin of the instantaneous
contributions in the case of a hard sphere fluid it is instructive
to write the shear viscosity \Ref{eta-total} in the standard form
\Ref{31} for conservative interactions, where the time correlation
function for $t>0$ is defined as,
\be\Label{51}
C_\eta(t)=2\eta_\infty \delta(t) +(\beta/ V)
\av{J_{-xy}e^{tL_+}J_{+xy}}_0,
\ee
and $\delta(t)$ is a Dirac delta function. It is then of interest
to compare this expression with the stress-stress correlation
function $C_\eta(t)$ in \Ref{31}, as recently calculated in
Ref.~\cite{DE04} for a conservative fluid with a repulsive power
law potential, $V(r)\sim r^{-\nu}$ (soft spheres) at short times at
large values of $\nu$. The short time dynamics has been calculated
exactly for the (very short) mean traversal time $\tau_\nu\sim
1/\nu$, required for particles to traverse the steep part of the
repulsive potential. It was shown that for asymptotically large
$\nu$, the short time behavior of $C_\eta (t)$ crosses over to a
contribution $\sim\delta(t)$, and agrees with the representation
\Ref{51}. This also demonstrates that smooth conservative
potentials $r^{-\nu}$ in the singular hard sphere limit as
$\nu\to\infty$, generate a $\delta(t)$-type short time correlation,
in agreement with the present results for hard spheres, which have
been derived from the very starting point for impulsive
interactions.

\section*{Acknowledgements}
M.H.E. is supported by Secretar\'\i a de Estado de Educaci\'on y
Universidades (Spain), and R.B. by the Universidad Complutense
(Profesores en el Extranjero). This work is financed by the
research project FIS2004-271 (Spain).

\setcounter{equation}{0}
\appendix

\section{ Hard sphere dynamics}
\subsection*{A1. Pseudo-Liouville and binary
collision operators}

The time evolution of any phase space function $A(t) = e^{t \calL}
A(0)$, evolving under the hard sphere dynamics \cite{EDHvL}, can be
described for positive times $(t>0)$ by the forward
pseudo-streaming operator, $\exp[tL_+]$, where $\calL =L_+$ is the
forward pseudo-Liouville operator for hard spheres. For backward or
time-reversed dynamics the pseudo-streaming operator is
$e^{t\calLinvert} = e^{-tL_-}\, (t>0)$. While true hard sphere
dynamics is undefined for physically inaccessible overlapping
configurations (with any $r_{ij}<\sigma$), the pseudo-streaming
operators are defined for all overlapping and non-overlapping
configurations. In the latter they generate the true hard sphere
dynamics, and in the former they can be defined conveniently.
Inside averages the pseudo-streaming operators are always preceded
by the hard sphere equilibrium distribution $\rho_0 \sim W_N
\exp[-\beta K]$, where $K$ is the total (conserved) kinetic energy.
The overlap function $W_N=1$ for non-overlapping configurations,
and $W_N=0$ for overlapping ones. Consequently, inside averages the
pseudo-streaming operators generate the correct forward and
backward hard sphere dynamics. For instance, the forward time
correlation function $\av{Ae^{tL_+}A}_0$ with $t>0$ is equal to the
time-reversed one $\av{Ae^{-tL_-}A}_0$.

In Sec.II we need, apart from the pseudo-Liouville operator
$\calL$, also its adjoint $\calLtilde$, its conjugate
$\calL^\epsilon$ with $\calLbar =\calLtilde^\epsilon$ according to
\Ref{14}. These operators can be expressed in binary collision
operators $T_{\pm}(ij)$ and $\overline T_{\pm}(ij)$ as follows,
\begin{eqnarray}\Label{A1}
\calL &=& \calL_0 +\sum_{i<j} T_+(ij)\equiv L_+ \nonumber \\
\calL^\epsilon &=& -  \calL_0 +\sum_{i<j} T_-(ij)\equiv -L_- \nonumber \\
\calLtilde &=& - \calL_0 +\sum_{i<j} \overline T_-(ij) \nonumber \\
\calLbar &=& \calL_0 +\sum_{i<j} \overline T_+(ij),
\end{eqnarray}
where $\calL_0= \sum_i \bv_i\cdot \partial/\partial\br_i$,
and the binary collision
operators satisfy the relations
\begin{equation}\Label{A2}
 \overline T_\pm =\widetilde T_\pm^\epsilon =\widetilde T_\mp ,
\end{equation}
as can be verified from the definitions below. The last equality on
the first and second line of (\ref{A1}) are the corresponding
notations used in the hard sphere kinetic theory (e.g.
\cite{dSEC-JSP,DE04}). The binary collision
operators are defined as
\begin{eqnarray}\Label{A3}
T_\pm(ij) & = & \sigma^{d-1} \int^{(\mp)} d\hat\bsigma
|\bv_{ij}\cdot\hat\bsigma|
\delta (\br_{ij}-\sigma\hat\bsigma)(b_\sigma - 1)\nonumber \\
\overline T_\pm(ij) & = & \sigma^{d-1} \int^{(\mp)} d\hat\bsigma
|\bv_{ij}\cdot\hat\bsigma| \left[\delta
(\br_{ij}-\sigma\hat\bsigma)b_\sigma - \delta
(\br_{ij}+\sigma\hat\bsigma)\right].
\end{eqnarray}
The superscript $(\mp)$ denotes the constraint $\mp
\bg_{ij}\cdot\hat\bsigma>0$, restricting the
$\hat\bsigma$-integration respectively to the pre-  and
post-collision hemisphere. The $b_\sigma$-operator is a
substitution operator, defined as,
\begin{eqnarray}\Label{A4}
b_\sigma\bv_i \equiv \bv_i' = \bv_i -
(\bv_{ij}\cdot\hat\bsigma)\hat\bsigma \nonumber \\
b_\sigma\bv_j \equiv \bv_j' = \bv_j +
(\bv_{ij}\cdot\hat\bsigma)\hat\bsigma  .
\end{eqnarray}
The definitions in \Ref{A1} imply that the pseudo-streaming
operator does {\it not} generate a dynamics, that is time-reversal
invariant, i.e. ${\cal L}^\epsilon \neq -{\cal L}$. On the other
hand the arguments above imply for hard sphere systems that $\av{A
B(t)}_0 = \av{A(-t) B}_0$ for $t>0$. Expressing this equality in
pseudo-streaming operators yields the first equality below, i.e.
\be \Label{A5}
\int d\Gamma \rho_0 A e^{t{\cal L}} B =\int d\Gamma \rho_0 B
e^{t{\cal L^\epsilon}} A = \int d\Gamma A e^{t\tilde{\cal
L}^\epsilon}  \rho_0 B.
\ee
In the last equality the adjoint of $\calL^\epsilon$ operator has
been introduced. The first equality is based on the properties
that $ \rho_0 \exp( t\calL ) = \rho_0 \exp( tL_+ ) $ with $t>0$,
and $ \exp( t\calL^\epsilon )\rho_0  = \exp(- tL_-) \rho_0 $ with
$t>0$ generate respectively the exact forward and backward
dynamics \cite{EDHvL}. So by comparing the utmost left and right
expressions we obtain  relation \Ref{16} for the hard sphere
streaming operators in the main text.

In the body of the paper we also need the property: If
$a_i=\{1,\bv_i,v_i^2\}$ is a collisional invariant, then
\begin{equation}\Label{A7}
T_\pm(ij)(a_i+a_j)=0.
\end{equation}

\subsection*{A2. Instantaneous transport coefficients
$\eta_\infty$ and $\lambda_\infty$}

Consider first $\Omega_{\eta\eta}$, which is defined as the
instantaneous shear viscosity $\eta_\infty$,
\begin{eqnarray}\Label{A8}
k^2\eta_\infty & = & -\rho\langle a_\bk^{\eta_i}|\calL
a_\bk^{\eta_i} \rangle \nonumber \\
&=& -( \beta/V) \langle g^*_{\bk y} (\calL_0 +\sum_{i<j}
T_+(ij))g_{\bk y}\rangle_0 ,
\end{eqnarray}
where (\ref{A1}), (\ref{215}) and (\ref{hydro-mode}) have been used
for $\calL$ and $a_k^{\eta _i}$, with $\hat\bk^i_\perp$ chosen
parallel to the $y$-axis. Moreover, $\calL_0$ gives a vanishing
contribution to (\ref{A8}). Calculation of the $T_+(ij)$
contribution for small $k$ requires the help of (\ref{A1}),
(\ref{A3}), (\ref{A7}) and (\ref{36}) to show that
\begin{equation}\Label{A9}
\eta_\infty  = -\frac{\beta m^2}{V}
\langle \sum_{i<j}( v_{iy} r_{ix} + v_{jy}r_{jx} )T_+(ij)
( v_{iy} r_{ix} + v_{jy}r_{jx} ) \rangle_0 .
\end{equation}
{From} here on the calculations for $\eta_\infty$ are identical to
those of (II.2)- (II.5) of Ref.\cite{DE04}, where the
Einstein-Helfand formulas for hard spheres were taken as a
starting point. We briefly sketch the different steps. For the
hard sphere fluid $\eta_\infty$ can be expressed in terms of the
hard sphere two-particle distribution function
$n^2\varphi_0(v_i)\varphi_0(v_j)\chi$, where $\chi =
g_0(r=\sigma+)$ is the radial distribution function at contact and
$\varphi_0(v) =(\beta m/\pi)^{d/2} \exp[-\frac{1}{2}
\beta m v^2]$ is the Maxwellian velocity distribution.

To carry out the subsequent integrations we change to the center of
mass, $\{\bR_{ij},\bG_{ij}=\frac{1}{2}(\bv_i+\bv_j)\}$, and
relative coordinates, $\{\br_{ij},\bv_{ij}=\bv_i-\bv_j\}$. The
result is
\begin{equation}\Label{A10}
\eta_\infty  = -\frac{1}{8} \beta \rho^2\chi
\int d\br \langle\langle g_x r_y T_+ (ij) g_x r_y\rangle\rangle,
\end{equation}
where $g_\alpha$ is the relative velocity, and $
\langle\langle\cdots\rangle\rangle $ denotes a
Maxwellian velocity average over all particles involved. Moreover
the $\br$-integration can be carried out because the operator $T_+$
contains $\delta^{(d)} (\br-\sigma\hat\bsigma)$.
 The remaining integrals are
$d$-dimensional generalizations of the collision integrals as
appearing in the Enskog theory for hard sphere fluids (see Chapter
16.8 of Ref.\cite{Chapman-Cowling}). Performing the $\hat\bsigma$-
and $\bg$-integrations yields finally for $\eta_\infty$ the result
listed in (\ref{33}) of the main text.

The calculations of the instantaneous heat conductivity
$\lambda_\infty$ runs completely parallel to that of $\eta_\infty$.
Starting from (\ref{215}) we obtain
\begin{eqnarray}\Label{A11}
k^2\lambda_\infty & = & -n {\cal C}_p \langle
a_\bk^T|\calL a_\bk^T\rangle \nonumber \\
& = & -\frac{\beta}{TV}
\langle (e^*_\bk-hn^*_\bk)
\sum_{i<j}T_+(ij)( e_\bk-hn_\bk) \rangle .
\end{eqnarray}
By performing the same steps  as for $\eta_\infty$,
one
sees that the terms involving the enthalpy
$hn_\bk$ do not contribute, and one arrives
at the result,
\begin{equation}\Label{A12}
\lambda_\infty = -\frac{\beta}{TV} \langle
\sum_{i<j} (e_ir_{ix} +e_j r_{jx}) T_+(ij)
(e_ir_{ix} +e_j r_{jx}) \rangle _0
\end{equation}
with $e_i =\frac{1}{2} mv_i^2$ for hard spheres. This expression is
completely analogous to (\ref{A9}), and a straightforward
evaluation leads to the expression (\ref{33}) for $\lambda_\infty$.

\end{document}